\documentclass[aps,prb,twocolumn,floatfix,floats,showpacs,superscriptaddress,raggedbottom]{revtex4}
\usepackage{graphicx,latexsym}
\usepackage{dcolumn}

\begin{document}

\title{Nonlinear Schroedinger-Poisson Theory for Quantum-Dot Helium}

\author{Gilbert Reinisch}
\email{Gilbert.Reinisch@oca.eu}
\affiliation{Universit\'e de Nice - Sophia Antipolis, CNRS, Observatoire de la C\^ote d'Azur,
             BP 4229, 06304 - Nice C\'edex 4, France}
\author{Vidar Gudmundsson}
\email{vidar@raunvis.hi.is}
\affiliation{Science Institute, University of Iceland, Dunhaga 3,
             IS-107 Reykjavik, Iceland}

%

\begin{abstract}
We use a nonlinear Schroedinger-Poisson equation to describe
two interacting electrons with opposite spins confined in a parabolic
potential, a quantum dot. We propose an effective form of the Poisson
equation taking into account the dimensional mismatch of the two-dimensional
electronic system and the three-dimensional electrostatics.
The results agree with earlier numerical calculations performed in a large
basis of two-body states and provide a simple model for continuous quantum-classical
transition with increasing nonlinearity. Specific intriguing properties due to
eigenstate non-orthogonality are emphasized.
\end{abstract}


\maketitle

%
%

\section{Introduction}
Quantum-dots can be viewed as artificially structured atoms in heterojunctions or
metal-oxide-semiconductor devices where few electrons are confined to a length
comparable to the mesoscopic effective Bohr radius $a_B$ ($a_B \sim 10^{-2}$
$\mu$m in the case of GaAs). Though the confinement can {\it a priori} occur in
all three directions, some types of experimentally realized quantum-dots display
an extension in the $x-y$ plane which is much larger than in  the growth
direction $z$ of the underlying semiconductor
structure.\cite{Sikorski89:2164a,Maksym90:108,Ashoori93:613} Therefore,
these quantum-dots are usually regarded as artificial atoms with a disk-like
shape. Since electron numbers $N$ as low as one or two per dot have already been
realized,\cite{Sikorski89:2164a,Ashoori93:613} quantum-dot Helium consisting of two electrons
trapped in the two-dimensional (2D) axisymmetrical harmonic potential
$V(r)= {1\over 2} M\omega^2r^2 $, where $r^2=x^2+y^2$ and $M$ is
the effective electron mass, is actually considered as the simplest realistic model
for an interacting quantum system.\cite{Merkt91:7320,Pfannkuche93:2244}
As itself or amongst other such few-electron systems, it has been
extensively studied in the relationship with the development
of nanotechnologies.\cite{Maksym96:10871,Wagner92:1951,Dineykhan97:13707,Hallam96:1452}
Both its exact 2D-3D analytical\cite{Dineykhan97:13707}
or 2D numerical\cite{Pfannkuche93:2244} solutions in the presence of a perpendicular
homogeneous external magnetic field are known, in particular by use of the
separation of the Hamiltonian into its center-of-mass and relative-motion terms,
due to the assumption of a parabolic particle confinement. Oscillations between
spin-singlet and spin-triplet ground states as a function of the magnetic field
strength have been predicted\cite{Wagner92:1951,Maksym90:108}
and experimentally observed.\cite{Ashoori93:613}
They are due to the interplay between the dot size and the strength of the magnetic
field. Another important competition occurs between the kinetic-energy matrix
elements and the electron-electron Coulomb interaction ones when changing the
characteristic length $L$ of the quantum dot without changing its shape.
Indeed, for small $L$, the Coulomb interaction becomes negligible and
the electrons behave like independent, uncorrelated particles.\cite{Bryant87:1140}
This happens in particular in the case of strong parabolic confinement
$\omega\rightarrow\infty$ since then $L\sim l_0=\sqrt{\hbar / M\omega}\rightarrow
0$.

In this rich theoretical and experimental context, we wish to
emphasize new physical results by use of a quite original -- with respect to the
above state of the art -- differential approach based on the Schroedinger-Poisson
(SP) definition of single-particle nonlinear eigenstates in quantum-dot Helium.
The problem with such eigenstates is that, being the (stationary) solutions of
the SP nonlinear differential system, they are \underbar{not} orthogonal. The
whole matrix machinery of quantum mechanics then fails and we are left to  return
to its {\it ab-initio} fundamental principles. In particular, the square scalar
product  ${\cal P}=\langle\Psi_a|\Psi_b\rangle^2\neq 0$ of two such nonlinear eigenstates
$\Psi_{a}$ and $\Psi_{b}$ defines the probability to find the system in either
state when it is known to be in the other one (this probability is of course zero
for orthogonal Hilbertian eigenstates). Equivalently -- and this
will be precisely shown below by use of the Fermi golden rule --, ${\cal P}$ yields
the transition probability either from $\Psi_{a}$ to $\Psi_{b}$ or reverse.
Therefore, if $\Psi_{a}$ is, say, the fundamental eigenstate and $\Psi_{b}$
is an excited one, ${\cal P}$ either measures the probability of an absorption
process ($\Psi_{a}\rightarrow\Psi_{b}$) or an emission one
($\Psi_{b}\rightarrow\Psi_{a}$). The following couple of questions are then
addressed and tentative answers provided: i) What energy is actually absorbed or
emitted as a consequence of the non-orthogonality of the nonlinear eigenstates
$\Psi_{a,b }$? ii) Are the affiliated energy exchanges  quantized and how?

The present paper is built as follows. We first display, by use of
standard numerical tools,\cite{Matlab:01} the remarkable properties of the solutions
$\{u_{\cal N},\,\,C_{\cal N} \}$ of the SP dimensionless nonlinear
differential system, where ${\cal N}$ is the nonlinear control parameter whose
value is given by the harmonic trap parabolicity $\omega$. Then we numerically
investigate the square scalar product ${\cal P}_{1\,3}=\langle u_1|u_3\rangle^2\neq 0$ of its
first two (zero-angular-momentum, for the sake of simplicity) eigenstates
$u_{1,3}$ when ${\cal N}$ is increased.
Subscripts always refer, in the present work, to those single-particle
energy eigenvalues (in units of $\hbar\omega$) which correspond to the ${\cal N}\rightarrow 0$
linear limit. We find an interference-like pattern $\propto \sin^2({1\over 2} {\cal N})$.
Then we validate the above results both by stressing the link between ${\cal P}_{1\,3}$
and Fermi's golden rule for small nonlinearity ${\cal N}\leq 1$, as well as
by displaying the transition of the system to the asymptotic semi-classical
Thomas-Fermi regime for high values of the nonlinearity ${\cal N}\gg 1$.
Finally, we test the reliability of the present SP description of quantum-dot Helium
by comparison with the existing numerical
(Refs (\onlinecite{Merkt91:7320,Pfannkuche93:2244}) \& (\onlinecite{Hallam96:1452}): respectively Figs
\ref{fig01}, \ref{fig03}, and \ref{fig05}) and analytical
(Ref.\ (\onlinecite{Dineykhan97:13707}): Fig.\ 1b) data concerning either the
fundamental energy level or, like in Ref.\ (\onlinecite{Hallam96:1452}), its (electro)chemical
potential. In all cases our nonlinear SP eigensolutions do agree surprisingly
well (i.e.\ within the percent) with the existing exact results (for comparison,
the Hartree and Hartree-Fock departures from the exact PGM fundamental energy
eigenvalue\cite{Pfannkuche93:2244} are respectively $44\,\,\%$ and $8\,\,\%$ in PGM's Fig.\ 3).

Therefore we recover an important property that has already been
emphasized in the $N=2$ Coulomb case, both for atomic Helium\cite{Reinisch01:042505} and
hydrogen ion $H^-$.\cite{Reinisch04:033613} Namely that the SP nonlinear differential description
yields surprisingly accurate values for the ground state energy when
compared to their corresponding mean-field Hartree-Fock ones.
The reason seems to be the very particular physical system that is actually
constituted by a mere Cooper-like pair of opposite-spin electrons trapped in the same orbital
bound state: one electron, say electron $a$ with orbital wave function $\Psi_a$, ``feels'' the
repulsive  electrostatic  potential $\Phi_b$ that is being created by its fellow
electron $b$ with orbital wave function $\Psi_b$. This potential $\Phi_b$ is defined by the
classical Poisson equation $\nabla\cdot\nabla \Phi_b\propto -|\Psi_b|^2$
while $\Psi_a$ is solution of the single-particle Schroedinger equation including both classical
potentials, namely the external confining potential $V(r)$ \underbar{and} the electrostatic
potential $\Phi_b$. The system is closed by the ``bosonic orbital assumption''
$\Psi_a\equiv \Psi_b$. As a consequence, there is no (positive) electron
self-interaction energy contribution like in Hartree's mean-field description.
Neither does (negative) Hartree-Fock's exchange energy play a significant role,
due to our $S=0$ opposite-spin assumption. Therefore, the only remaining
difference with respect to the exact corresponding energy eigenvalues might be
due -- or at least partially -- to the next-order (negative) correlation
effects.

%
\section{The 2D radial Schroedinger-Poisson nonlinear ordinary differential system}
The SP differential system results from coupling the
single-particle stationary Schroedinger equation that defines the 2D
orbital wave function $\Psi(x,y)$ in the potential $V(x,y) + \Phi(x,y) $
\begin{equation}
      \Bigl[-{\hbar^2\nabla^2\over 2M} + V(x,y) +
      \Phi(x,y) \Bigr]\Psi(x,y)=\mu \Psi(x,y),
\label{Eq01}
\end{equation}
with the Poisson equation which solely defines the mutual electrostatic repulsive
interaction $\Phi(x,y) $ between the two particles\cite{Reinisch01:042505,Reinisch04:033613}
\begin{equation}
      \nabla^2 \Phi(x,y)=-2\pi {\cal N}\hbar\omega\,|\Psi(x,y)|^2.
\label{Eq02}
\end{equation}
Since $|\Psi|^2 \propto [{\rm length}]^{-2}$, we \underbar{must} indeed introduce, for
dimensional reasons related to Eq.\ (\ref{Eq02}), a characteristic energy which we wish to write as
${1\over 2}{\cal N}\hbar\omega$. The corresponding dimensionless parameter
${\cal N}\equiv {\cal N}(\omega)$ will be defined below and is a typical measure of the
SP nonlinearity. It is important to keep in mind that the above system
Eqs (\ref{Eq01}-\ref{Eq02}) is only relevant for particles in the same orbital state $\Psi$.

Assuming the 2D axisymmetrical parabolic confining potential
$V(x,y)=V(r) = {1\over 2} M\omega^2 r^2$, we have:
\begin{equation}
      \Psi(x,y)=\Psi(r,\,\phi) =  \psi(r)\,e^{\displaystyle im\phi}.
\label{Eq03}
\end{equation}
The wavefunction $\Psi$ is thus the eigenstate of the angular-momentum operator
$-i\hbar\partial / \partial\phi$ related to its eigenvalue $m\hbar$.
Its radial part $\psi(r)$, which describes the 2D confinement of the electron
system in the $z=0$ transverse plane with radial symmetry in agreement with
the experimental results, is defined by
\begin{equation}
      {\ddot u}+{1\over X}{\dot u} + \Bigl[C-{X^2\over 4} \Bigr]u=0;\quad
      \quad {\ddot C}+{1\over X}\,{\dot C} + {4 m^2\over X^4}= u^2 ,
\label{Eq04}
\end{equation}
if we introduce the following dimensionless quantities ($l_0=\sqrt{\hbar/M\omega}$
is the characteristic parabolic length)
\begin{equation}
      X=\sqrt{2}\,{r\over l_0};\quad
      u=\sqrt{\pi\hbar {\cal N}\over M\omega}\psi  ;\quad
      C={\tilde \mu}-{m^2\over X^2}-{\tilde\Phi},
\label{Eq05}
\end{equation}
in order to scale the Poisson nonlinearity (namely, the r.h.s.\ of Eq.\ (\ref{Eq02}) )
to unity, as evidenced by the r.h.s.\ of Eq.\ (\ref{Eq04}b). The dot stands for derivation
with respect to the (dimensionless) radius $X$ and, as already emphasized, the tilde
superscript labels energy in units of $\hbar\omega$.
The single-particle probability of presence $|\Psi|^2$ must be normalized to unity.
Therefore ${\cal N}$ is the corresponding dimensionless norm of the solution $u$
\begin{equation}
      \int\,|\Psi|^2\,d^2{\bf x}=1\quad\rightarrow\quad
      \int_0^{\infty}u^2\, XdX= {\cal N}.
\label{Eq06}
\end{equation}
%

\section{The nonlinear quantum-classical transition}
\begin{figure}[htbq]
      \includegraphics[width=0.49\textwidth,angle=0]{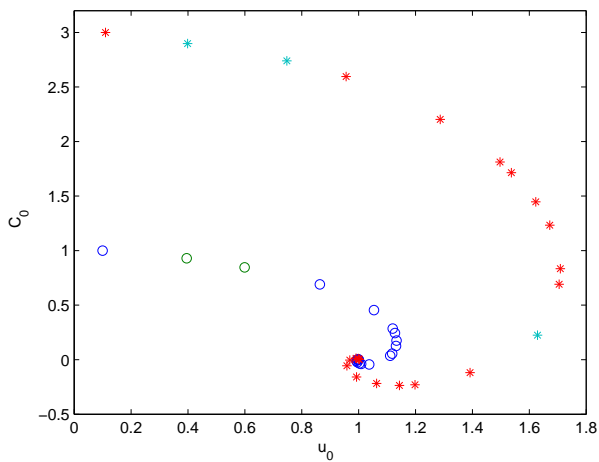}
      \caption{(Color online). The convergence of the discrete SP nonlinear system towards
              the quasi-classical continuum Thomas-Fermi regime defined by the fixed point
              $u \equiv 1$ and $C_0={1\over 4}[X^2]_{X=0}=0$ when ${\cal N}$ increases from
              $\sim 10^{-2}$ to $\sim 10^2$ in the $\{C_0$ vs $u_0\}$ boundary condition phase
              space for the two first $m=0$ nonlinear eigenmodes defined by their corresponding
              linear quantum numbers: namely (cf.\ Eq.\ (\ref{Eq07c})) $n=0,\,m=0$ ($u_1$: circles) and
              $n=1,\,m=0$ ($u_3$: stars).}
\label{fig01}
\end{figure}

The dimensionless solution of our differential
problem Eqs (\ref{Eq04}) yields $u(r)$ and $C(r)$ as functionals of ${\cal N}$ given by
Eq.\ (\ref{Eq06}). It is defined by the initial conditions $u_0=u(0)$, ${\dot
u}_0={\dot u(0)}$, $C_0=C(0)$ and ${\dot C}_0={\dot C(0)}$. Amongst them, $u_0$
and $C_0$ are left free and will be chosen by numerical dichotomy, in order to
yield regular bound-state eigenstates, i.e.\ $u$ solutions defined by $u(X)\rightarrow 0$ for
$X\rightarrow\infty$ (for practical purposes, it will be sufficient to impose
$u(X) < 10^{-7}$  at $X=9$: see Fig.\ \ref{fig02} below). On the other hand, ${\dot C}_0=0$
(no potential cusp at $X=0$) and the determination of the last remaining parameters,
namely ${\dot u}_0$, proceeds from step-by-step  increase of both $u_0$ and ${\dot u}_0$,
starting from their linear limit where $u_0\sim\sqrt{{\cal N}}\ll 1$
\begin{equation}
      u_0 \ll 1 ; \quad  {\dot u}_0\sim {\dot u}^{\rm lin}(0).
\label{Eq07a}
\end{equation}
Here
\begin{equation}
         u^{\rm lin}(X)\propto e^{-{X^2\over 4}}\,X^{m\over 2}\,P_n(X),
\label{Eq07b}
\end{equation}
defines\cite{Pfannkuche93:2244} the ${\cal N}\rightarrow 0$ linear solutions in terms of the quantum
numbers $n$ and $m$ and of the Laguerre polynomial $P_{n}(X)$, namely $P_0=1$;
$P_1=1-{1\over 2}X^2$; $P_2=2-2X^2+{1\over 4}X^4$; $P_3=6-9X^2+{9\over 4}X^4-{1\over 8}X^6$ ...
These single-particle linear parabolic states correspond to the energy eigenvalues
\begin{equation}
      E^{\rm lin}=E_{n,m}=(2n+|m|+1)\,\hbar\omega .
\label{Eq07c}
\end{equation}
Figure \ref{fig01} displays the $C_0$ versus $u_0$ spiraling trajectories for the two first
$m=0$ eigenstates  $u_1$ (circles) and $u_3$ (stars).
Recall that the subscripts always refer
in the present work to the number of $\hbar\omega$ quanta present in the ${\cal
N} \rightarrow 0$ linear limit of the single-particle energy, in accordance with
Eq.\ \ref{Eq07c}. These trajectories in the $\{u_0,\,\,C_0\}$ plane are parametrized
with respect to increasing values of the nonlinear parameter ${\cal N}$, i.e.\
with decreasing values of the trap harmonicity $\omega$. Indeed electron-electron
interaction becomes relatively (with respect to quantum kinetic energy)
more and more important when the two electrons are
less and less confined (see above Part I).
Actually  ${\cal N}$ varies in Fig.\ \ref{fig01} from $10^{-2}$
($u_0\sim 0.1$) to $10^{2}$ ($u_0\sim 1$) where one then reaches the
quasi-classical asymptotic Thomas-Fermi regime. This regime is defined by
neglecting the  quantum kinetic derivative terms in Eq.\ (\ref{Eq04}a), thus yielding
$C(X)\sim X^2/4$ and hence $C_0=C(0)=0$, while $u(X)\equiv 1$ through Eq.\ (\ref{Eq04}b).
Therefore the initial conditions for the two  discrete modes $u_1$ and $u_3$
converge towards the Thomas-Fermi fixed point $\{u_0=1;\,C_0=0\}$ for ${\cal
N}\rightarrow\infty$ as evidenced by Fig.\ \ref{fig01}.
Physically, this means that there is a continuous transition, through the increase
of nonlinearity in the system, from the  ${\cal N} \leq \pi$ ``pure'' quantum regime where
the  quantum kinetic energy defined by the derivative terms in Eq.\ (\ref{Eq04}a) plays a  major
role towards the ${\cal N} \gg \pi$ classical one  where the dimensionless
Schroedinger equation Eq.\ (\ref{Eq04}a) reduces to its last-bracket classical-energy
term. As a consequence, the ${\cal N} \rightarrow\infty$ highly nonlinear case
leads to the progressive merging of the two discrete energy levels $u_1$ and $u_3$
into the single one whose initial conditions are defined by the fixed point
displayed in Fig.\ \ref{fig01}. Therefore quantum eigenstate discreteness disappears, which
is the hallmark of the classical regime: a continuous energy spectrum
sets on about the uniform wavefunction profile $u(X)\equiv 1$ and $C(X)\equiv 0$, where
the chemical potential equals the -- here vanishing, due to our $m=0$ assumption
-- centrifugal potential  \underbar{plus} the electrostatic interaction
potential, as shown by Eq.\ (\ref{Eq05}c).\hfill\break The onset of the first
corresponding oscillation in the amplitude of the respective modes $u_1(X)$
(continuous line) and $u_3(X)$ (dotted line) is displayed in Fig.\ \ref{fig02}.
\begin{figure}[htbq]
      \includegraphics[width=0.48\textwidth,angle=0]{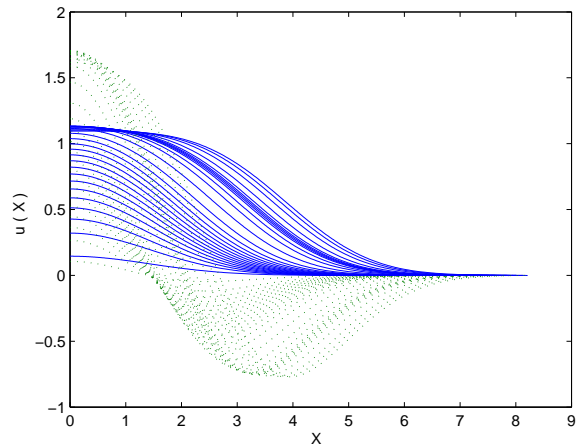}
      \caption{(Color online). Several nonlinear eigenstate profiles $u_1(X)$ (continuous line)
              and $u_3(X)$ (dashed  line) for increasing  values of the dimensionless quantum-dot
              size $k$, namely $10^{-2} \leq k \leq 8.7$, where $k=l_0/a^*$ is the ratio of
              the characteristic harmonic length $l_0=\sqrt{\hbar/M\omega}$ over the effective
              Bohr radius $a^*=\hbar^2/Me^2$. The maximum-amplitude thresholds at $u_0 \sim 1.2$
              (resp.\ $u_0 \sim 1.8$) for the ground-state mode $u_1$ (resp.\ the excited mode $u_3$)
              and displayed by Fig.\ \ref{fig01} are clearly visible (profile accumulation effect).}
      \label{fig02}
\end{figure}

\section{Chemical potential and energy: the explicit definition ${\cal N}(\omega)$}
In the following, we shall only consider zero-angular-momentum $m=0$ states for
the sake of simplicity (we have indeed checked that $m\neq 0$ nonlinear eigenstates are
equally well described by the above differential system: see below Fig.\ \ref{fig04} where the
quantum-dot spectra  are  displayed versus their corresponding ${\cal N}$ for $n\leq 3;\,\,m\leq 3$).
The SP virial energy $E$ per particle corresponding to the nonlinear
eigenstate $u(X)$ is twice the expectation value  $\langle{1\over 2}\,M\omega^2
r^2\rangle$  of the external parabolic potential energy (virial theorem for a harmonic potentiel).
In terms of the dimensionless quantities
defined in Eqs (\ref{Eq05}), it reads
\begin{equation}
      {\tilde E}={1\over 2{\cal N}}\,\int_0^{\infty}u^2\,X^3 dX .
\label{Eq08}
\end{equation}
On the other hand, the chemical potential $\mu$ defined by Eq.\ (\ref{Eq01}) is that
energy which is required in order to add the second  electron to the single-electron
quantum-dot (Koopman's theorem\cite{Szabo96:01}). It can truly be regarded as the nonlinear
eigenvalue of the SP differential system related to the corresponding nonlinear eigenstate
$\Psi$ (or $u$ in the reduced units defined in Eqs (\ref{Eq05})). Therefore we have
\begin{equation}
      {\tilde \mu} =2{\tilde E}-{\tilde E }^{\rm lin} ,
\label{Eq09}
\end{equation}
where $E^{\rm lin}=E_{n,m} $ is defined by Eq.\ \ref{Eq07c}.
In the present work where we only consider $m=0$ eigenstates, the two first levels
are ${\tilde E_1}$ (resp.\ ${\tilde E_3}$) corresponding to $n=0$ (resp.\  $n=1$) in
units of $\hbar\omega$. The nonlinear integrodifferential system Eqs
(\ref{Eq04}-\ref{Eq09}) is closed by  the use of Eq.\ (\ref{Eq05}c)  at $X=0$. This yields the (reduced)
chemical potential ${\tilde\mu}=\mu/\hbar\omega$ for $m=0$
\begin{eqnarray}
       {\tilde \mu}=C(0)+{\tilde\Phi}(0)&=&C_0+{e^2\over \hbar\omega }\,
       \int\,{|\Psi|^2\over r}\,d^3{\bf x}\nonumber\\ &=&C_0+{\sqrt{2}\,k\over{\cal N}}\,
       \int_0^{\infty} u^2\,dX ,
\label{Eq10}
\end{eqnarray}
where $k=l_0/a^*$ is the usual dimensionless dot size corresponding to the harmonic
length $l_0=\sqrt{\hbar/M\omega }$ and the effective Bohr radius $a^*=\hbar^2/Me^2$
(ranging from $a^*=67\,\,nm$ for InSb to $a^*=9.8\,\,nm$ for GaAs).

\begin{figure}[htbq]
      \includegraphics[width=0.48\textwidth,angle=0]{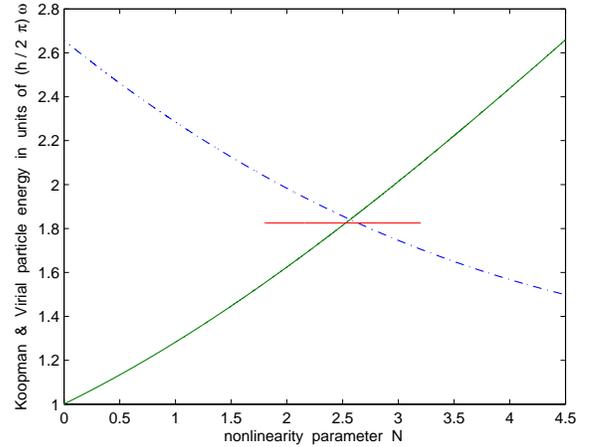}
      \caption{(Color online). For GaAs quantum-dot Parahelium defined by the confinement
              $\hbar\omega=3.37$ meV,  the  dielectric constant of the bulk material
              $\kappa =12.4$ and the effective mass $M=0.067\, m_e$ (where $m_e$ is
              the electron mass),\cite{Pfannkuche93:2244} we have ${\cal N} =2.53$ and ${\tilde  E}=1.831$ by
              use of (respectively) Eqs (\ref{Eq11a}) \& (\ref{Eq11b}). This last SP ground-state
              energy per particle value appears here as the intersection of its virial (continuous line)
              and its Koopman (dashed-dotted line) definitions as respectively provided by Eqs
              (\ref{Eq08}) \& (\ref{Eq09}-\ref{Eq10}). PGM's exact numerical value
              ${\tilde  E}={1\over 2}(12.28)$ meV $/3.37$
              meV = 1.822 given in Ref.\ (5) is plotted as the horizontal segment.}
      \label{fig03}
\end{figure}

\begin{figure}[htbq]
      \includegraphics[width=0.48\textwidth,angle=0]{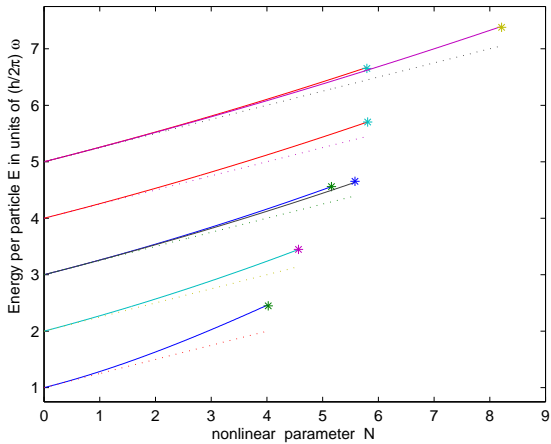}
      \caption{(Color online).
              Illustration of Eq.\ (\ref{Eq11c}) (dotted line) for the
              $1\leq {\tilde E}^{\rm lin}\leq 5$ nonlinear eigenmodes (as defined by their linear
              ${\cal N}\rightarrow 0$ quantum numbers $n$ \& $m$: see Eq.\ (\ref{Eq07c})).
              The stars display the greatest $u_0$ value reached for each mode: see Fig.\
              1 for the two first $m=0$ modes where the stars would respectively
              correspond to $u_0 \sim 1.2$ and $u_0 \sim 1.7$.
              From bottom to top (a \& b superscripts label degeneracies which are
              lifted by nonlinearity): $\{n=0,\,m=0\}$, $\{n=0,\,m=1\}$, $\{n=1,\,m=0\}^a$, $\{n=0,\,m=2\}^a$,
              $\{n=1,\,m=1\}$, $\{n=2,\,m=0\}^b$, $\{n=1,\,m=2\}^b$.}
      \label{fig04}
\end{figure}

Equations (\ref{Eq04}-\ref{Eq10}) self-consistently define, for any given value of the trap
characteristic harmonic frequency $\omega$ (or, equivalently, its reduced size $k$),
the solution $u\equiv u_{\omega}(X)$, its norm ${\cal N}\equiv {\cal N}({\omega})$
as well as its corresponding single-particle energy ${\tilde E}\equiv {\tilde E}({\omega})$,
together with the chemical potential ${\tilde \mu}\equiv {\tilde\mu}({\omega})$.
We numerically obtain, for instance for the ground state in the ``quantum-regime''
interval of values ${\cal N}\leq \pi$ (see below)
\begin{equation}
      {\cal N}(k)\sim  {0.8839\, k\over 0.4218 + 0.1247\, k}
\label{Eq11a}
\end{equation}
while its energy is
\begin{equation}
      {\tilde E}({\cal N})\sim  1+0.24670\, {\cal N}+0.03683\,
      {\cal N}^2 - 0.00217\, {\cal N}^3 .
\label{Eq11b}
\end{equation}
Therefore $ {\cal N}(k)$ is a monotonic increasing function of the dot size,
starting like ${\cal N}\sim 2k$ for small values of the dot size $k$, while
\begin{equation}
        {\tilde E}\sim {\tilde E^{\rm lin}}+{1\over 4}{\cal N} \quad @
        \quad {\cal N}< 1 \quad\leftrightarrow\quad k< {1\over 2} .
\label{Eq11c}
\end{equation}
In Ref.\ \onlinecite{Pfannkuche93:2244}, for instance, where $\hbar\omega=3.37$ meV for a GaAs parabolic quantum dot
($M$ equals $0.067$ electron mass while the charge is $1/\sqrt{12.4}$ electron
charge), we have $a^*=9.79$ nm and $l_0=18.5$ nm. Hence $k=1.89$. Then Eqs
(\ref{Eq11a},\ref{Eq11b}) respectively yield ${\cal N}=2.53$ and ${\tilde E}={\tilde E}_{{\rm
per\,particle}}=1.83$. Therefore $E_{{\rm quantum\,dot}}= 2 {\tilde E}_{{\rm
per\,particle}} = 2(1.83)\,\hbar\omega=12.33$ meV to be compared with PGM's value $12.28$ meV: see
Fig.\ \ref{fig03} where the virial energy per particle (solid line), defined by Eq.\ (\ref{Eq08}),
is plotted together with the Koopman one (dashed line), defined from Eqs (\ref{Eq07c}-\ref{Eq10}).
The energy per particle ${\tilde E}$ is defined
by the intersection of both plots.
Similarly, in Ref.\ \onlinecite{Hallam96:1452}, $\hbar\omega=2$ meV yields $k=2.43$.
Hence ${\cal N}=2.97$ and ${\tilde E}=2.00$, which yields ${\tilde \mu}=2{\tilde
E}-1=3.00$ in accordance with Eq.\ (\ref{Eq09}), and therefore $\mu=3.00\,\hbar\omega=6.00$
meV which is in complete agreement with the Coulomb-interaction case ($d_1=d_2=\infty$) of that
reference.

A remarkable property of the SP nonlinear eigensolutions is their ``universal'' limit
behavior defined by Eq.\ (\ref{Eq11c}) for small ${\cal N}$, whatever the actual state's quantum
numbers $n$ and $m$ in Eq.\ (\ref{Eq07c}): see Fig.\ \ref{fig04} where
${\tilde E}= {\tilde E^{\rm lin}}+{1\over 4}{\cal N} $
is plotted in dashed line.
Therefore the energy
\begin{equation}
      \Delta = {1\over 2}{\cal N} \hbar\omega ,
\label{Eq12a}
\end{equation}
which was introduced for dimensional reasons into the Poisson equation (\ref{Eq02})
is simply the smallest additional quantum-dot energy due to particle-particle
interaction nonlinearity. Indeed Eq.\ (\ref{Eq11c}) yields for the $2E$ quantum-dot energy
\begin{equation}
      \lim_{{\cal N}\rightarrow 0}2E=2  E^{\rm lin}+\Delta ,
\label{Eq12b}
\end{equation}
where $E^{\rm lin}$ is the ${\cal N}=0$ linear energy per particle defined by Eq.\ (\ref{Eq07c}).
Therefore $\Delta$ is the smallest interaction (or nonlinear) energy value in our
two-electron SP  system that comes in addition to the already existing ``linear''
quanta $\hbar\omega$, when ${\cal N}\rightarrow 0$. To see whether $\Delta$ is a
true ``nonlinear quantum'' of energy -- i.e.\ whether the energy exchanges
between the two levels $E_1$ and $E_3$ can be described in the terms of both
$\hbar\omega$ \underbar{and} $\Delta$ -- demands to define the actual transition
probability between these levels from the non-orthogonality of the corresponding
eigenstates. This will be done in the next part.

\section{The scalar product $\langle u_1|u_3\rangle$ and the corresponding transition
         probability}
Let us define the (normalized) scalar product
\begin{equation}
      \langle u_1|u_3\rangle ={1\over \sqrt{{\cal N}_1{\cal N}_3}}\,
      \int_0^{\infty}u_1u_3\,XdX ,
\label{Eq13a}
\end{equation}
together with
\begin{equation}
      {\displaystyle  \int_0^{\infty}\,u_1^2 X^3 \,dX - {\cal N}_1
      \bigl(1+C_0^{(1)}\bigr)
      \over \displaystyle  \int_0^{\infty}\,u_3^2 X^3 \,
      dX -{\cal N}_3\bigl(3+C_0^{(3)}\bigr)} =
      {\displaystyle \int_0^{\infty}\,u_1^2 \,  dX \over \displaystyle
      \int_0^{\infty}\,u_3^2 \, dX} ,
\label{Eq13b}
\end{equation}
of both zero-angular-momentum eigenstates $u_1(X)$ and $u_3(X)$
corresponding to those energies $E_1$ and $E_3$ which
are respectively defined at ${\cal N}_{1,3}\rightarrow 0$ by $E_{0,0}$ and
$E_{1,0}$ in Eq.\ (\ref{Eq07c}). Equation (\ref{Eq13b}) is the matching condition
\begin{equation}
      k(\omega) =  {{\cal N}_1\bigl({\tilde\mu}_1-C_0^{(1)}\bigr)\over
      \sqrt{2}\int_0^{\infty}\,u_1^2 \,  dX} =   {{\cal
      N}_3\bigl({\tilde\mu}_3-C_0^{(3)}\bigr)\over \sqrt{2}\int_0^{\infty}\,
      u_3^2 \,  dX} ,
\label{Eq13c}
\end{equation}
defined by Eqs (\ref{Eq07a}-\ref{Eq10}) which states that the trap harmonicity $\omega$, or equivalently
its quantum-dot dimensionless length $k=\sqrt{\hbar/M\omega }/(\hbar^2/Me^2)$,
must be identical for the two modes $u_{1,\,3}$ that enter the calculation of the scalar
product (\ref{Eq13a}). Practically, in the numerical simulations of Eqs (\ref{Eq13a}-\ref{Eq13c}),
we will consider Eq.\ (\ref{Eq13b}) as verified if it is fulfilled within an $10^{-6}$ error.

\begin{figure}[htbq]
      \includegraphics[width=0.48\textwidth,angle=0]{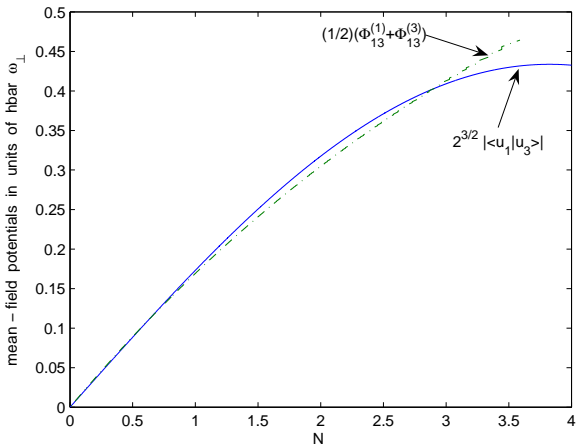}
      \caption{(Color online). the convergence for small  nonlinearity
              ${\cal N}\sim {\cal N}_1\sim {\cal N}_3 < 1$ of the SP
              nonlinear model towards Fermi's ``golden rule'' probability transition defined by
              Eqs (\ref{Eq13a}-\ref{Eq14b}), as displayed by Eq.\ (\ref{Eq16}).}
      \label{fig05}
\end{figure}

In order for the scalar product Eqs (\ref{Eq13a}-\ref{Eq13c}) to make physical sense, we wish
to link it with standard time-independent linear perturbation theory in the case of
small nonlinearity ${\cal N} \ll  1$, i.e.\ for small ``perturbative''
particle-particle interaction $\Phi$ defined by Eqs (\ref{Eq01}-\ref{Eq02}). Therefore we deduct that
\begin{equation}
      \lim_{{\cal N}\rightarrow 0}\langle u_1|u_3\rangle ^2 ={\cal P}_{1\,3} ,
\label{Eq14a}
\end{equation}
where
\begin{equation}
      {\cal P}_{1\,3}= {4\over\hbar^2}\bigr|
      \langle u^{\rm lin}_3 |H^{\rm pert} |u^{\rm
      lin}_1\rangle \bigl|^2\,\,\,{\sin^2 ({1\over 2}
      \omega_{1\,3}\,t) \over\omega _{1\,3}^2} ,
\label{Eq14b}
\end{equation}
together with $\hbar\omega_{1\,3}=E_3-E_1\sim 2 \hbar\omega$, yields
Fermi golden rule's transition probability per particle.\cite{Feynman65:01} Indeed both the energies
$E_{i}\sim E_{{1\over 2}(i-1),0}$ per particle ($i=1,3$: cf.\ Eq.\ (\ref{Eq07c})) and the
corresponding normalized eigenstates $u_i\sim u^{\rm lin}_i$ in Eq.\ (\ref{Eq14b}) are
those corresponding to the \underbar{unperturbed} linear system (hence the superscript),
namely\cite{Pfannkuche93:2244}  $u^{\rm lin}_1=e^{-X^2/4}$ and $u^{\rm lin}_3=(1-{X^2\over 2})e^{-X^2/4}$
(cf.\ Eqs (\ref{Eq07b})). The perturbation
potential ${\tilde H^{\rm pert}}$  is equal to ${1\over 2}{\bar{\tilde\Phi}}$
(per particle: hence the factor ${1\over 2}$) where ${\bar{\tilde\Phi}}={1\over
2}\bigl[{\tilde\Phi}^{(1)}+{\tilde\Phi}^{(3)}\bigr]$ is the interaction potential
that has been averaged over its two components ${\tilde\Phi}^{(1)}$ and
${\tilde\Phi}^{(3)}$. Therefore
\begin{eqnarray}
      \langle u^{\rm lin}_3 |{\tilde H^{\rm pert}} |u^{\rm lin}_1\rangle =
      {\tilde H^{\rm pert}}_{1\,3}&=&
      {1\over 2} \bigl[{\bar{\tilde\Phi}}_{1\,3}\bigr]\nonumber\\ &=&
      {1\over 4}\bigl[{\tilde\Phi}^{(1)}_{1\,3}+
      {\tilde\Phi}^{(3)}_{1\,3}\bigr] ,
\label{Eq15}
\end{eqnarray}
where the matrix elements ${\tilde\Phi}^{(i)}_{1\,3}$ ($i=1,3$) have
been calculated by use of the above-mentioned linear normalized eigenstates $u^{\rm lin}_{1,3}$.
In the stationary perturbative regime where $\omega_{1\,3}\,t\sim 2\omega t \gg 1\gg
H^{\rm pert} t/\hbar$, the time-dependent term in Eq.\ (\ref{Eq14b}) can be replaced  by
its averaged value ${1\over 2}$, yielding ${\cal P}_{1\,3}={1\over
2}\bigl[{\tilde H^{\rm pert}}_{1\,3}\bigr]^2= {1\over
32}\bigl[{\tilde\Phi}^{(1)}_{1\,3}+{\tilde\Phi}^{(3)}_{1\,3}\bigr]^2$.
Therefore Eq.\ (\ref{Eq14a}) becomes
\begin{equation}
      \lim_{{\cal N}\rightarrow 0}{\bar{\tilde\Phi}}_{1\,3}={1\over 2}
      \lim_{{\cal N}\rightarrow 0}
      \bigl[{\tilde\Phi}^{(1)}_{1\,3}+{\tilde\Phi}^{(3)}_{1\,3}\bigr] =
      2\sqrt{2}\,|\langle u_1|u_3\rangle | .
\label{Eq16}
\end{equation}
Figure \ref{fig05} displays the r.h.s.\ of Eq.\ (\ref{Eq16}) (continuous line) versus its l.h.s.\
(dashed-dotted line) and shows the numerical fulfillement of this condition.

Therefore the square scalar product defined by Eqs (\ref{Eq13a}-\ref{Eq13c}) yields, in the limit of
small nonlinearity ${\cal N}_{1,3}$, the transition probability ${\cal P}_{1\,3}$
from the fundamental nonlinear eigenstate $u_1$ to the excited one $u_3$ or
reverse. On the other hand, we showed in Section III that the quasi-classical
Thomas-Fermi regime yields $\lim_{{\cal N}\rightarrow \infty}u_{1,3}\equiv 1$
(see Fig.\ \ref{fig01}). The two modes $u_{1,3}$ then become equivalent. Consequently the
transition probability between them should obviously become equal to unity, which
is consistent with $[\langle u_1|u_3\rangle^2]_{u_1\sim u_3}=1$ from definition
(\ref{Eq13a}). Therefore it seems quite natural to extrapolate to \underbar{all} values
of the nonlinearity ${\cal N}_{1,3}$ the physical meaning of
$\langle u_1|u_3\rangle ^2$ in terms of the transition probability ${\cal P}_{1\,3}$
as defined by Eqs (\ref{Eq14a}-\ref{Eq14b}).

\section{Quantum transitions between two $m=0$ nonlinear eigenstates}
Let us now proceed to the investigation of the quantum transitions
between the two nonlinear eigenstates $u_{1,3}$ by use of the numerical calculation
of the scalar product defined by Eqs (\ref{Eq13a}-\ref{Eq13c}).
It consists in increasing the nonlinearity through a
three-loop iterative scheme from the ${\cal N}\ll 1$ linear regime. The two first
loops define each eigenstate $u_{1,3}$ which vanish with a $10^{-7}$ accuracy at $X\sim 9$
which is our numerical value for $X\sim\infty$ (see Fig.\ \ref{fig02}) while the third one
evaluates the matching condition Eq.\ (\ref{Eq13b}) within $10^{-6}$ and then calculates
the scalar product given by Eq.\ (\ref{Eq13a}). The integrals which appear in
Eqs (\ref{Eq13a}-\ref{Eq13c})
are transformed into additional first-order ordinary differential equations with
vanishing initial conditions whose solutions  are  taken at $X\sim 9$. Then the
whole resulting differential system is numerically integrated by use of standard
tools.\cite{Matlab:01}

\begin{figure}[htbq]
      \includegraphics[width=0.48\textwidth,angle=0]{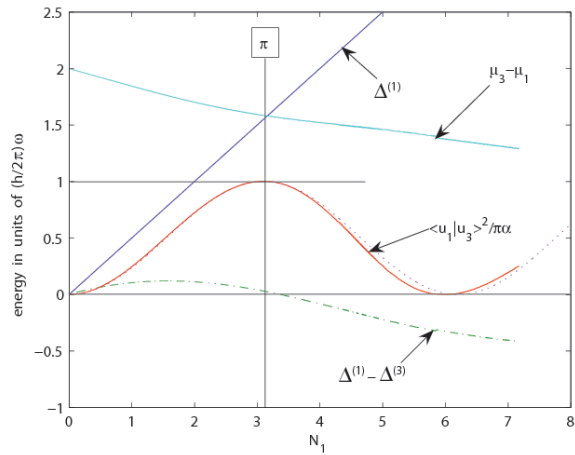}
      \caption{(Color online). Equation (\ref{Eq17})'s square scalar product
              $\langle u_1|u_3\rangle^2/\pi\alpha$
              normalized by $\pi\alpha=\pi/137.036$, as compared with its $\sin^2$ approximation
              (dotted line), together with the nonlinear resonance condition
              Eq.\ (\ref{Eq18a}) (in ``tilde''
              units of $\hbar\omega$): i)  intersection of the two upper plots that yields
              ${\tilde \mu}_3-{\tilde \mu}_1\sim {\tilde \Delta}_1\sim\pi/2$, and ii) intersection of
              the dashed-dotted lower plot with zero that yields
              ${\tilde \Delta}_1\sim {\tilde \Delta}_3$,
              i.e.\ ${\cal N}_{1}\sim {\cal N}_{3}$, at the resonance nonlinearity ${\cal N}_1\sim\pi$.}
      \label{fig06}
\end{figure}

Figure \ref{fig06} displays the following remarkable interference-like pattern with respect to
the ground-state nonlinear parameter ${\cal N}_1$.
Intriguing enough, since
the present SP differential model is non-relativistic (there is no velocity of light in it),
it is best scaled  by use of the numerical value of the fine-structure-constant
$\alpha=e^2/\hbar c$ multiplied by $\pi$, namely $\pi/137.036=2.2925...\,\,10^{-2}$
\begin{equation}
      {1\over \pi \alpha}\langle u_1|u_3\rangle^2 \sim 1.0005  \,\sin^2\bigl[0.5060{\cal N}_1\bigr] .
\label{Eq17}
\end{equation}
We note that, when ${\cal N}_1 > 4$, the departure from the r.h.s.\ of Eq.\ (\ref{Eq17})
(dotted line in Fig.\ \ref{fig06}) becomes significant as the transition toward the
asymptotic quasi-classical Thomas-Fermi regime sets on. On the other hand, Fig.\ \ref{fig06}
displays the following chemical-potential gap transition process
\begin{equation}
      \mu_3- \mu_1 \sim {\Delta}_{\pi} ,
\label{Eq18a}
\end{equation}
where ${\Delta}_{\pi}\sim {\Delta_1}\sim {\Delta_3}\sim {\displaystyle{\pi\over
2}}\hbar\omega_{\pi}$ is the common characteristic energy $\Delta$,
defined by Eq.\ (\ref{Eq12a}),  of the two eigenstates $u_{1,\,3}$ about the maximum
(of amplitude $1.0005\,\pi\alpha$) of their square scalar product
$\langle u_1|u_3\rangle^2$, i.e.\ at the very peculiar quantum-dot nonlinearity
${\cal N}_1 \sim {\cal N}_3 \sim\pi$ related to the particular
$\omega=\omega_{\pi}$ trap parabolicity. This value corresponds to
the specific parabolic confinement $\hbar\omega_{\pi}  \sim 0.14\,\epsilon$
where $\epsilon=Me^4/\hbar^2$ is the effective quantum-dot's atomic energy unit:
$\epsilon=  11.86$ meV for AsGa, thus yielding $\hbar\omega_{\pi} \sim 1.66$ meV
and $\Delta_{\pi}\sim 2.61$ meV while $\epsilon=  27.21$ eV if the dielectric
constant of the bulk material equals unity, then yielding $\hbar\omega_{\pi} \sim
3.80$ eV and therefore $\Delta_{\pi}\sim 5.97$ eV. According to Eq.\ (\ref{Eq09}),  Eq.\
(\ref{Eq18a}) yields the corresponding quantization rule for the $2E$ quantum-dot energy
at $\omega\sim \omega_{\pi}$
\begin{equation}
      2(E_3-E_1) \sim 2\hbar\omega_{\pi}+ {\Delta}_{\pi} .
\label{Eq18b}
\end{equation}
Equations (\ref{Eq12b}) and (\ref{Eq18b}) show that the characteristic energy ${\Delta}$
which scales the electrostatic particle-particle interaction through the
nonlinear differential Poisson equation (\ref{Eq02}) is in fact a true ``nonlinear quantum''.
Indeed, on the one hand, it is the smallest particle-particle interaction energy present
in the system at vanishing nonlinearity ${\cal N}\rightarrow 0$.
On the other hand, the maximum of the $\langle u_1|u_3\rangle^2$ transition
probability between the two states is reached at resonance, i.e.\ either when ${\Delta}$ equals their
nonlinear-eigenvalue chemical-potential gap or when their quantum-dot energy gap is but the
mere sum of the two standard ``linear'' radial quanta $\hbar\omega$ \underbar{and} ${\Delta}$.

\section{Conclusion and perspectives}
In the present paper, we have described the parabolic quantum dot by use of the nonlinear
differential eigenproblem Eqs (\ref{Eq01}-\ref{Eq02}) and emphasized its relevance with respect to all
existing corresponding results in the literature. This  Schroedinger-Poisson (SP) differential
system yields new quantum concepts such as the non-orthogonal nonlinear eigenstates $\Psi$ and
their corresponding chemical-potential nonlinear eigenvalues $\mu$.\cite{Reinisch07:120402}
In order to comply with the dimensional self-consistence between the
two-dimensional electronic system and its three-dimensional electrostatics, we scaled the Poisson
equation according to the characteristic energy
$\Delta={1\over 2}{\cal N}\hbar\omega$ where ${\cal N}$ is a normalized (see Eq.\ (\ref{Eq06})) measure of
the system nonlinearity. We showed that $\Delta$ is actually the
true ``nonlinear energy quantum'' of the system for:
i) it is  the smallest additional ``nonlinear'' particle-particle interaction energy with respect
to the standard ``linear'' radial harmonic quantum $\hbar\omega$  when ${\cal N}\rightarrow 0$
(see Eq.\ (\ref{Eq12b}));
ii) it fits with that nonlinear-eigenvalue (or chemical-potential) gap between the two first
zero-angular-momentum eigenstates  which occurs about the maximum of their square scalar product
$\langle u_1|u_3\rangle^2$ (see Eqs (\ref{Eq18a}-\ref{Eq18b})), i.e.\  about the maximum of their transition
probability ${\cal P}$ (as a consequence of Fermi's golden rule).

Further developments of the present work should (non exhaustively)
address the two following experimental, numerical as well as theoretical topics:

1) Could the nonlinear resonance defined by Eqs (\ref{Eq18a}-\ref{Eq18b}) at the very particular trap
parabolicity $\hbar\omega=\hbar\omega_{\pi}$ ($=1.66$ meV for GaAS) be observable
and how? It would be a definite plus for the present model to provide an
opportunity for experimental verification.

2) The $\pi\alpha$ scaling adopted in Eq.\ (\ref{Eq17}) seems extremely accurate.
Indeed the square scalar product maximum divided by $\pi$ approaches the numerical value
$1/137.036$ of the fine-structure-constant $e^2/\hbar c$ within $0.05\,\%$ in the
latest state of our numerical simulations
\begin{equation}
      {1\over\pi}\langle u_1|u_3\rangle^2_{max}={1\over 136.97} .
\label{Eq19}
\end{equation}
It seems hard to believe that Eq.\ (\ref{Eq19}) is but the result of a mere numerical coincidence.
Rather, we wish to point out that Eq.\ (\ref{Eq19}) might echo Feynman's emphasis of such a ``magic
number''.\cite{QED:01}  This stunning non-relativistic property of the nonlinear-eigenstate
square-scalar-product scaling will be further investigated in a future publication.

\begin{acknowledgments}
      The authors acknowledge financial support from the Research
      and Instruments Funds of the Icelandic State,
      the Research Fund of the University of Iceland.
      One of the authors (GR) feels indebted to
      P.\ Valiron for sharp criticisms and very useful
      discussions and comments. He thanks J.\ Bec for double-checking the
      numerical procedures.
\end{acknowledgments}
%
%

%
%
\bibliographystyle{apsrev}

%
%
%
\end{document}